\documentclass[smallextended]{svjour3}       
\smartqed  
\usepackage{graphicx}
\usepackage{amsmath,bbm,amssymb,multirow,times,bm}
\usepackage{color}
\usepackage{enumitem}
\usepackage{mathtools}
\usepackage{tikz}
\usepackage{url}
\usepackage{changepage}
\usepackage{hyperref}
\usepackage[ruled]{algorithm2e}
\usepackage{subfigure}
\journalname{Quantum Information Processing}

\begin{document}

\title{Quantum locally linear embedding for nonlinear dimensionality reduction\thanks{This work is supported by the National Key R\&D Program of China, Grant No. 2018YFA0306703.}}



\author{Xi He \and
		Li Sun \and
		Chufan Lyu \and
		Xiaoting Wang
}


\institute{Xi He \at
              Institute of Fundamental and Frontier Sciences, University of Electronic Science and Technology of China, Chengdu, Sichuan, 610054, China \\
              \email{xihe@std.uestc.edu.cn} 
\and
            Li Sun \and Chufan Lyu \and Xiaoting Wang \at
              Institute of Fundamental and Frontier Sciences, University of Electronic Science and Technology of China, Chengdu, Sichuan, 610054, China \\
}

\date{Received: date / Accepted: date}

\maketitle

\begin{abstract}
Reducing the dimension of nonlinear data is crucial in data processing and visualization. The locally linear embedding algorithm (LLE) is specifically a representative nonlinear dimensionality reduction method with well maintaining the original manifold structure. In this paper, we present two implementations of the quantum locally linear embedding algorithm (QLLE) to perform the nonlinear dimensionality reduction on quantum devices. One implementation, the linear-algebra-based QLLE algorithm, utilizes quantum linear algebra subroutines to reduce the dimension of the given data. The other implementation, the variational quantum locally linear embedding algorithm (VQLLE) utilizes a variational hybrid quantum-classical procedure to acquire the low-dimensional data. The classical LLE algorithm requires polynomial time complexity of $N$, where $N$ is the global number of the original high-dimensional data. Compared with the classical LLE, the linear-algebra-based QLLE achieves quadratic speedup in the number and dimension of the given data. The VQLLE can be implemented on the near term quantum devices in two different designs. In addition, the numerical experiments are presented to demonstrate that the two implementations in our work can achieve the procedure of locally linear embedding.

\end{abstract}

\section{Introduction}
\label{sec:introduction}
The dimensionality reduction in the field of machine learning refers to using some methods to map the data points in the original high-dimensional space into some low-dimensional space. The reason why we reduce the data dimension is that the original high-dimensional data contains redundant information, which sometimes causes errors in practical applications. By reducing the data dimension, we hope to reduce the error caused by the noise information and to improve the accuracy of identification or other applications. In addition, we also want to find the intrinsic structure of the data through the dimensionality reduction in some cases. Among the many dimensionality reduction algorithms, one of the most widely used algorithms is the principal component analysis (PCA). PCA is a linear dimensionality reduction algorithm embedding the given data into a linear low-dimensional subspace~\cite{1,2}. In addition to the PCA algorithm, the linear discriminant analysis algorithm (LDA)~\cite{3} and the A-optimal projection algorithm (AOP)~\cite{4} both show outstanding performance in reducing the data dimension.

However, all the algorithms mentioned above actually belong to linear dimensionality reduction techniques. It means that they are not so efficient in dealing with nonlinear data to some extent. Thus, some nonlinear dimensionality reduction algorithms are proposed to give special treatment to nonlinear high-dimensional data. The locally linear embedding algorithm (LLE) which was proposed in 2000 by Sam T.Roweis and Lawrence K.Saul is typically one of the most representative nonlinear dimensionality reduction algorithms~\cite{5}. It can preserve the original topological structure of the data set during the process of dimensionality reduction. At present, LLE has been widely applied in all kinds of fields such as data visualization, pattern recognition and so on.  

Although classical algorithms can accomplish the machine learning tasks effectively, quantum computing techniques can be applied to the realm of machine learning resulting in quantum speedup. Based on the quantum phase estimation algorithm~\cite{6}, quantum algorithm for linear systems of equations~\cite{7}, Grover's algorithm~\cite{8} and so on, all kinds of quantum machine learning algorithms have been proposed. Representatively, quantum support vector machine~\cite{9}, quantum data fitting~\cite{10} and quantum linear regression~\cite{11} were proposed respectively to deal with problems of pattern classification and prediction. Quantum deep learning algorithms such as quantum Boltzmann machine~\cite{12}, quantum generative adversarial learning~\cite{13,14} and so on were also put forward to exhibit their capabilities in quantum physics. In addition, there are some variational quantum-classical hybrid algorithms presented for machine learning recently~\cite{15,16,17}. In general, quantum machine learning has developed into a vibrant interdisciplinary field.     

In the field of quantum dimensionality reduction, there are also some outstanding techniques. The quantum principal component analysis algorithm (qPCA) was proposed with exponential speedup compared with the classical PCA algorithm~\cite{18}. Afterwards, the quantum linear discriminant analysis (qLDA) was designed for dimensionality reduction and classification~\cite{19}. Recently, the quantum A-optimal projection algorithm (QAOP) presented in~\cite{20} performs superior regression performance. Different from all the above-mentioned algorithms, in this paper, we present two implementations of the quantum locally linear embedding algorithm (QLLE) for nonlinear dimensionality reduction. Compared with the classical LLE algorithm, we can invoke the quantum $k$-NN algorithm to find out the $k$ nearest neighbors of all the given data with quadratic speedup in the data preprocessing stage~\cite{21}. In the main part of the QLLE algorithm, the linear-algebra-based QLLE algorithm can be implemented in $O(\mathrm{poly}(\log N))$, which achieves exponential speedup compared with the classical LLE algorithm in time $O(\mathrm{poly}(N))$ where $N$ is the number of the original high-dimensional data points. The VQLLE algorithm can be performed on the near term quantum devices with a variational hybrid quantum-classical procedure. Specifically, two different designs of the VQLLE are presented, namely the end-to-end VQLLE and the matrix-multiplication-based VQLLE. In addition, the numerical experiments of the two implementations are presented demonstrating that the QLLE algorithms in our work can achieve the procedure of the locally linear embedding. 

In our work, the contributions are mainly reflected in two aspects. One contribution, two implementations of the QLLE are presented for nonlinear dimensionality reduction. The linear-algebra-based QLLE can be performed on a universal quantum computer with quantum speedup. The variational QLLE can be implemented on near term quantum devices without the requirement of fully coherent evolution. The other contribution, we design the corresponding quantum circuits and numerical experiments for the QLLE algorithms making the execution of our theoretical analysis realizable.

In section~\ref{sec:classical_LLE}, the classical LLE is briefly overviewed. Subsequently, the linear-algebra-based QLLE is presented in section~\ref{sec:LABQLLE}. In addition, the implementation of the VQLLE is described in section~\ref{sec:VQLLE}. To evaluate the feasibility and performance of the QLLE algorithms, the numerical experiments are provided in section~\ref{sec:numerical experiments}. Finally, we make a conclusion and discuss some open questions in section~\ref{sec:conclusion}. 

\section{Classical locally linear embedding}
\label{sec:classical_LLE}
\begin{figure}
\centering
\includegraphics[width=0.75\textwidth]{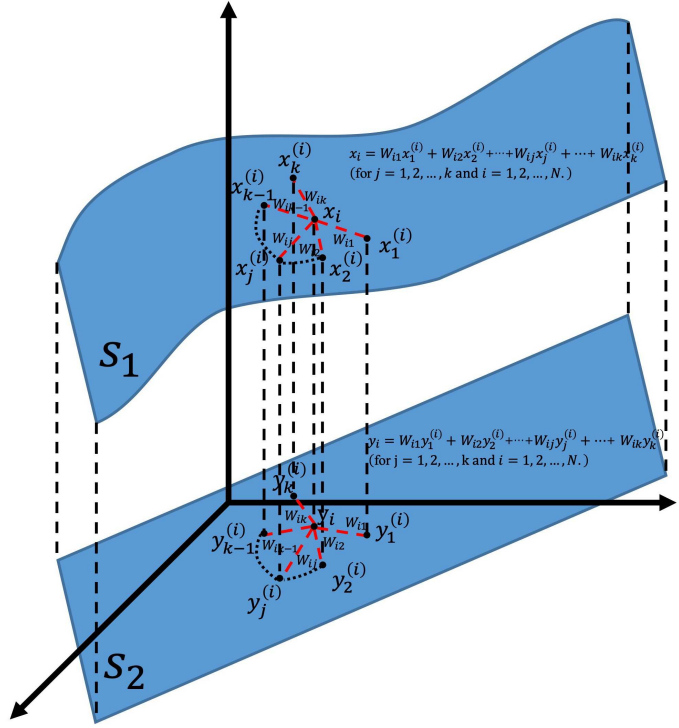}
\caption{The schematic diagram of the classical LLE algorithm. The high-dimensional data are distributed in a manifold $S_{1}$ and each data point $x_{i}$ is assumed to be represented by the linear combination of its $k$ nearest neighbors $x^{(i)}_{j}$. The LLE algorithm aims to find the low-dimensional data $y_{i}$ embedded in $S_{2}$ with keeping this linear relationship, which means that the weight matrix $W$ is invariable during the dimensionality reduction}
\label{fig:LLE}
\end{figure}

In this section, the classical LLE algorithm is described in detail~\cite{5}. The LLE algorithm aims to reduce the dimension of the high-dimensional data by embedding them from the original high-dimensional space to some low-dimensional space. During the procedure of dimensionality reduction, the linear relationships between the data points and their corresponding $k$ nearest neighbors remain unchanged. The schematic diagram of the classical LLE algorithm is presented in Fig.~\ref{fig:LLE}.

The overall procedure of the LLE can be summarized as follows:

(1) Find out the $k$ nearest neighbors of each high-dimensional data point $x_{i}$ with $i = 1, 2, \dots, N$.

(2) Construct the local reconstruction weight matrix $W$ with the $k$ nearest neighbors of $x_{i}$.

(3) Compute the low-dimensional data $Y$ with the reconstruction weight matrix $W$.

Herein, we review the classical LLE algorithm in detail. Suppose we have the input data set $X = \{x_i \in \mathbb{R}^D: 1\leq i \leq N\}$. LLE is a local dimensionality reduction algorithm which mainly utilizes the local linearity among the data points to approximate their global property. Thus, in the first place, LLE attempts to find out the $k$ nearest neighbors of each data point with the $k$-NN algorithm to subsequently construct the reconstruction weight matrix $W$. As a matter of fact, $W$ is the intermediary that connects the entire dimensionality reduction procedure. Having acquired the $k$ nearest neighbors of all the data points, we can then set the cost function
\begin{equation}\label{eq:phi_w}
    \begin{split}
        &\min_W \varPhi(W) = \sum_{i=1}^N \left \vert x_i - \sum_{j=1}^k W_{ij}x_j^{(i)} \right \vert^2 \\
        &\quad s.t. \sum_{j=1}^k W_{ij} = 1, 
    \end{split}
\end{equation}
where $x_j^{(i)}$ represents the $j$th nearest neighbor of $x_{i}$, $W_{ij}$ denotes the weight coefficient of $x_{j}^{(i)}$.

Equivalently, the matrix form of the cost function $\varPhi(W)$ is
\begin{equation}\label{eq:matrix_phi_W}
	\begin{split}
		&\min_W \varPhi(W) = \sum_{i=1}^N W_i^T (\Delta X_{i})^T (\Delta X_{i})W_i = \sum_{i=1}^{N}W_{i}C_{i}W_{i}, \\
    	&\quad s.t. W_{i}^{T}\textbf{1}_{N} = 1,
	\end{split}
\end{equation}
where $W_i=(W_{i1},W_{i2},\dots,W_{ik},0,\dots,0)^T \in \mathbb{R}^{N}$; $\Delta X_{i} = [(x_i-x_1^{(i)}),\dots,(x_i-x_k^{(i)}), 0,\dots,0] \in \mathbb{R}^{D \times N}$; $C_i = (\Delta X_{i})^T (\Delta X_{i})$ and $\textbf{1}_{N} = (1, 1, \dots, 1)^{T}_{N}$ is a $N$-dimensional vector with all the elements equaling $1$.

Applying the method of Lagrangian multiplier on Eq.~\eqref{eq:matrix_phi_W}, we can obtain
\begin{equation}\label{eq:W_{i}}
    W_i = \frac{C_i^{-1} \textbf{1}_{N}}{\textbf{1}_{N}^T C_i^{-1} \textbf{1}_{N}},
\end{equation}
and the specific derivation of $W_{i}$ is presented in Appendix~\ref{A}.

In practice, we can efficiently get the $W_i$ by solving the linear system of equations $C_i W_i = \textbf{1}_{N}$, and then rescaling the weights for normalization. Hence, the reconstruction weight matrix $W = (W_{1}, W_{2}, \dots, W_{i}, \dots, W_{N})_{N \times N}$ can be subsequently constructed.

Assuming the low-dimensional data set after dimensionality reduction is $Y = \{y_i \in \mathbb{R}^d: 1\leq i \leq N \}$, we want to keep the linear relationship during the process of dimensionality reduction. It is equivalent to minimizing the cost function
\begin{equation}\label{eq:phi_y}
    \begin{split}
        \min_Y \ &\varPhi(Y) = \sum_{i=1}^N \left \vert y_i - \sum_{j=1}^k W_{ij} y_j^{(i)} \right \vert^2, \\
        s.t. &\sum_{i=1}^N y_i = 0; \\
        &\frac{1}{N}\sum_{i=1}^N y_i y_i^T = I_d,
    \end{split}
\end{equation}
where $y_j^{(i)}$ represents the $j$th nearest neighbor of $y_{i}$; $W_{ij}$ denotes the weight coefficient of $y_{j}^{(i)}$ and is exactly the same as the weight coefficient in Eq.~\eqref{eq:phi_w}.

Similarly, Eq.~\eqref{eq:phi_y} can be transformed to its matrix form
\begin{equation}\label{eq:matrix_phi_y}
    \begin{split}
            \min_Y \ &\varPhi(Y) = \Vert Y - Y W \Vert_F^2, \\
    s.t.\ &Y \textbf{1}_{N} = \textbf{0};  \\
    &\frac{1}{N}YY^T = I_d,  
    \end{split}
\end{equation}
where $\Vert . \Vert_{F}$ represents the Frobenius norm.

Having set $\varPhi(Y)$ properly, we can subsequently transform Eq.~\eqref{eq:matrix_phi_y} into
\begin{equation}\label{eq:final_phi_y}
    \min \varPhi(Y) = \mathrm{tr}(YMY^{T}),
\end{equation}
where the target matrix $M = (I_N - W)(I_N - W^T)$. The derivation with Lagrangian multipliers is presented in Appendix~\ref{B}.

Therefore, the problem of minimizing $\varPhi(Y)$ can be transformed into solving the $2$nd to the $(d+1)$th smallest eigenvectors of $M$ (the first eigenvalue of $M$ is 0, and the corresponding eigenvector is $\textbf{1}_N$, so it does not reflect the data characteristics). Finally, we get the low-dimensional data set $Y = (y_1,y_2,...,y_N) = (u_2,u_3,...,u_{d+1})^T \in \mathbb{R}^{d \times N}$, where $u_2,u_3,...,u_{d+1}$ stand for the $2$nd to the $(d+1)$th eigenvectors of $M$ which are corresponding to the relative smallest eigenvalues in ascending order.

\section{Linear-algebra-based quantum locally linear embedding}
\label{sec:LABQLLE}
In this section, the implementation of the linear-algebra-based QLLE is presented. In the data preprocessing, we invoke the quantum $k$-NN algorithm to find the $k$ nearest neighbors of the given data. Subsequently, the low-dimensional data $Y$ can be obtained by utilizing the quantum basic linear algebra subroutines. 

\subsection{Data preprocessing}
\label{subsec:data_preprocessing}
To present the QLLE algorithm, the input data vector $x_i = (x_{1i},x_{2i},...,x_{Di})^T$ can be encoded as a $q$-qubit quantum state $|x_i\rangle$ where $q = \log D$. Assume that the quantum states $| x^{(i)}_{j} \rangle$ for $j = 1, \dots, k$ represent the $k$ nearest neighbors of $| x_{i} \rangle$. With invoking the quantum $k$-NN algorithm, we can find out the $k$ nearest neighbors $\{| x^{(i)}_{j} \rangle: 1 \leq j \leq k \}$ of $| x_{i} \rangle$ for $i = 1, 2, \dots, N$. Compared with the classical $k$-NN algorithm which should construct the corresponding data structure index in $O(N \log N)$ and search the $k$ nearest neighbors in $O(k \log N)$, the quantum $k$-NN algorithm~\cite{21} can be implemented with quadratic speedup in $O(R \sqrt{kN})$ where $R$ is the times of Oracle execution.

\subsection{Construction of the weight matrix $W$}
\label{subsec:construction_of_W}
By the procedure of data preprocessing, all the original quantum states and their corresponding $k$ nearest neighbors can be obtained. Subsequently, we present how to construct the weight matrix $W$. 

In the first place, we prepare the quantum states $| x_{i} - x^{(i)}_{j} \rangle$ with the data points $x_{i}$ and the corresponding nearest neighbors $x^{(i)}_{j}$ where $i = 1, 2, \dots, N$ and $j = 1, 2, \dots, k$. It is note that the elements which are not corresponding to the $k$ nearest neighbors are not under consideration according to section~\ref{sec:classical_LLE}. With knowing all the elements of $x_{i}$, the corresponding $q$-qubit quantum state $| x_{i} \rangle = | a_{q} a_{q-1} \dots a_{1} \rangle$. We apply the quantum Fourier transform (QFT)~\cite{22} on $| x_{i} \rangle$ resulting in 
\begin{equation}\label{eq:psi_x_{i}}
	| a_{q} a_{q-1} \dots a_{1} \rangle \xrightarrow{\textbf{QFT}} | \phi_{q}(a) \rangle \otimes | \phi_{q-1}(a) \rangle \otimes \dots \otimes | \phi_{p}(a) \rangle \otimes \dots \otimes | \phi_{1}(a) \rangle,
\end{equation}
where $| \phi_{p}(a) \rangle = \frac{1}{\sqrt{2}} (| 0 \rangle + e^{2 \pi i 0.a_{p}a_{p-1} \dots a_{1}} | 1 \rangle)$ for $p = 1, 2, \dots, q$.

\begin{figure}
\centering
\includegraphics[width=\textwidth]{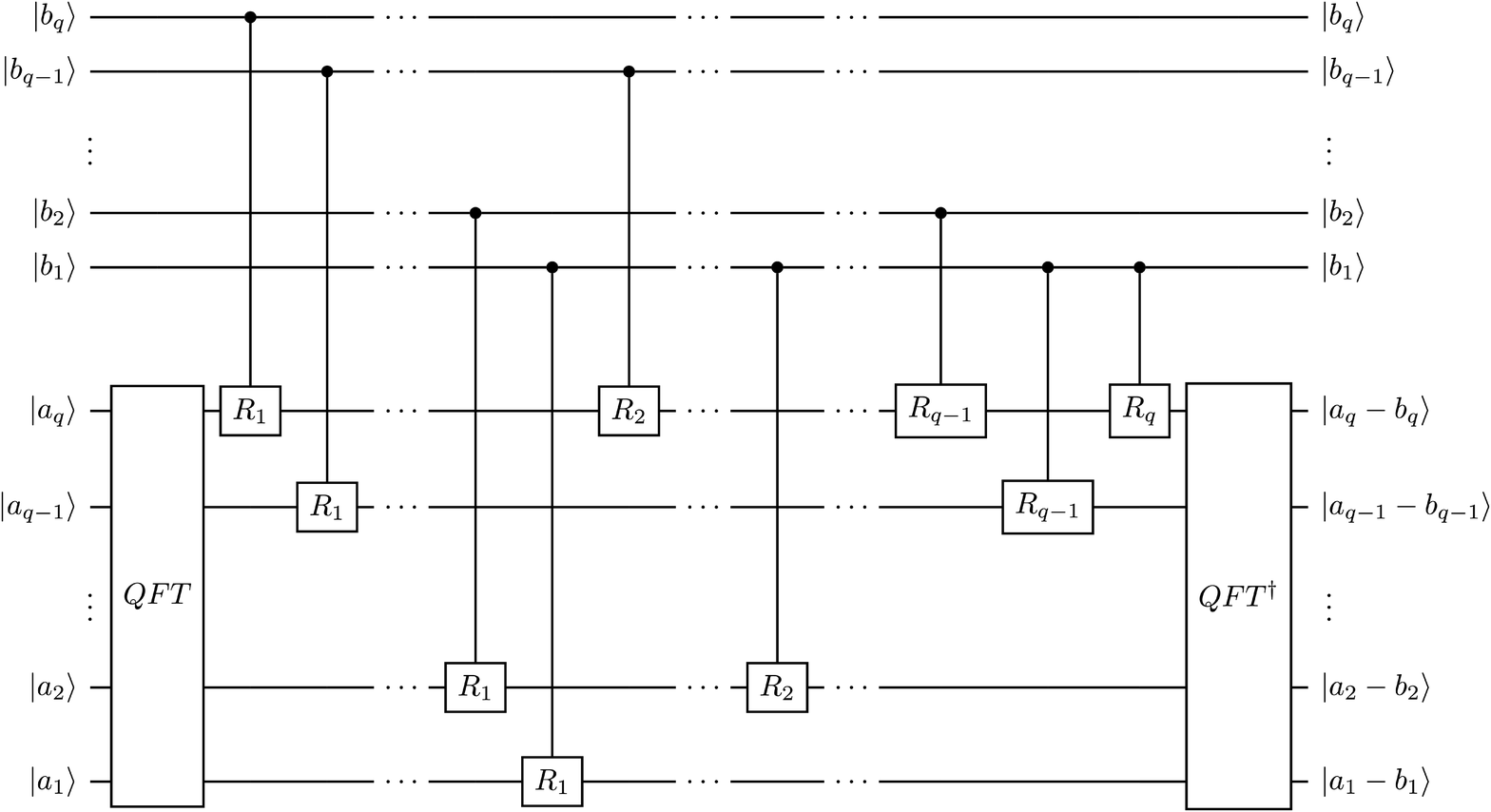}
\caption{The quantum circuit of the quantum subtractor computing $| x_{i} - x^{(i)}_{j} \rangle$}
\label{fig:subtractor}
\end{figure}

Similarly, $x^{(i)}_{j}$ can be encoded as $| x^{(i)}_{j} \rangle = | b_{q} b_{q-1} \dots b_{1} \rangle$. Subsequently, we perform the controlled rotation operation $\textbf{U}_{\textbf{P}} = | 0 \rangle \langle 0 | \otimes I + | 1 \rangle \langle 1 | \otimes R_{P}$ as shown in Fig.~\ref{fig:subtractor} to prepare the quantum state $| \phi_{q}(a-b) \rangle \otimes | \phi_{q-1}(a-b) \rangle \otimes \dots \otimes | \phi_{1}(a-b) \rangle$, where $R_{P}= \begin{bmatrix}
	1 & 0 \\
	0 & e^{-2 \pi i / 2^{p}}
\end{bmatrix}$ with $p = 1, 2, \dots, q$. It is note that the controlled rotation operation of the quantum subtractor we adopt can be performed in parallel with $q \log q$ operations~\cite{23,24,25}. After performing the inverse QFT on $| \phi_{q}(a-b) \rangle \otimes | \phi_{q-1}(a-b) \rangle \otimes \dots \otimes | \phi_{1}(a-b) \rangle$, the quantum state $| x_{i} - x^{(i)}_{j} \rangle = | a_{q} - b_{q} \rangle \otimes | a_{q-1} - b_{q-1} \rangle \otimes \dots \otimes | a_{1} - b_{1} \rangle$ can be obtained. By iteratively invoking the quantum subtractor, we can finally acquire all the quantum states $| x_{i} - x^{(i)}_{j} \rangle$ with $j = 1, 2, \dots, k$. The quantum circuit of the quantum subtractor is presented in Fig.~\ref{fig:subtractor}.

\begin{figure}
\centering
\includegraphics[width=0.5\textwidth]{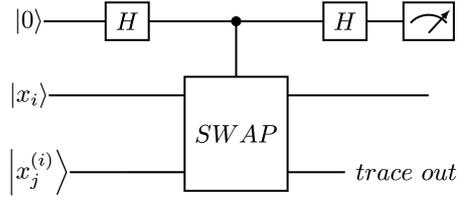}
\caption{The quantum circuit of computing the inner product $\langle x_{i} | x^{(i)}_{j} \rangle$}
\label{fig:swap_test}
\end{figure}

In addition to solving the quantum states $| x_{i} - x^{(i)}_{j} \rangle$, we also need to compute the norms $\left | x_{i} - x^{(i)}_{j} \right |$. It is obvious that
\begin{equation}\label{eq:norm_x_{i}-x^{(i)}_{j}}
	\left | x_{i} - x^{(i)}_{j} \right |^{2} = | x_{i} |^{2} + | x^{(i)}_{j} |^{2} - 2 \textrm{Re}\langle x_{i}, x^{(i)}_{j} \rangle.
\end{equation} 
With the given data and the procedure of data preprocessing, the norms $| x_{i} |$ and $| x^{(i)}_{j} |$ are trivial. As to the third term of Eq.~\eqref{eq:norm_x_{i}-x^{(i)}_{j}}, we firstly prepare the initial state 
\begin{equation}\label{eq:psi_0}
	| \psi_{0} \rangle = \frac{1}{\sqrt{2}} (| 0 \rangle | x_{i} \rangle + | 1 \rangle | x^{(i)}_{j} \rangle). 
\end{equation}
Then, the Hadamard gate is performed on the ancilla to obtain the state
\begin{equation}\label{eq:psi_{1}}
	| \psi_{1} \rangle = \frac{1}{2} \left [ | 0 \rangle (| x_{i} \rangle + | x^{(i)}_{j} \rangle) + | 1 \rangle (| x_{i} \rangle - | x^{(i)}_{j} \rangle) \right ].
\end{equation}
We measure the ancilla resulting in the probability of $| 0 \rangle$
\begin{align}\label{eq:P(0)}
	P_{1}(0) &= \frac{1}{4} (\langle x_{i} | + \langle x^{(i)}_{j} |) (| x_{i} \rangle + | x^{(i)}_{j} \rangle) \notag \\
	&= \frac{1}{4} (2 + \langle x_{i} | x^{(i)}_{j} \rangle + \langle x^{(i)}_{j} | x_{i} \rangle) \notag \\
	&= \frac{1}{2} + \frac{1}{2} \textrm{Re} \langle x_{i} | x^{(i)}_{j} \rangle.
\end{align}
Thus, the third term of Eq.~\eqref{eq:norm_x_{i}-x^{(i)}_{j}}
\begin{align}\label{eq:third_term}
	2 \textrm{Re} \langle x_{i}, x^{(i)}_{j} \rangle &= 2 | x_{i} | | x^{(i)}_{j} | \langle x_{i} | x^{(i)}_{j} \rangle \notag \\
	&= (4P_{1}(0) - 2) | x_{i} | | x^{(i)}_{j} |.
\end{align}
In practice, we achieve this procedure with the swap test~\cite{26,27} circuit as shown in Fig.~\ref{fig:swap_test}. The input state is $| 0 \rangle | x_{i} \rangle | x^{(i)}_{j} \rangle$ and we trace out the third register before the measurement to acquire the state $| \psi_{1} \rangle$. Finally, we can compute the inner product with the probability of measuring $| 0 \rangle$ on the ancilla qubit.

Having access to $|x_{i} - x^{(i)}_{j} \rangle$ and $| x_{i} - x^{(i)}_{j} |$, we can utilize the quantum random access memory (qRAM) to construct the state
\begin{equation}\label{eq:psi_{delta_x_i}}
	| \psi_{\Delta \tilde{X}_{i}} \rangle = \frac{1}{\sqrt{ \sum_{j=1}^{k} \sum_{m=1}^{D} | x_{mi} - x^{(i)}_{mj} |^{2}}} \sum_{j=1}^{k} \sum_{m=1}^{D} ( x_{mi} - x^{(i)}_{mj} ) | j \rangle | m \rangle 
\end{equation}
where $\Delta \tilde{X}_{i} = [ (x_{i} - x^{(i)}_{1}), \dots, (x_{i} - x^{(i)}_{k}) ]$. Afterwards, we trace out the $| m \rangle$ register resulting in the density operator
\begin{align}\label{eq:rho_{tilde_C_{i}}}
	\rho_{\tilde{C}_{i}} &= \mathrm{tr}_{m} \{| \psi_{\Delta \tilde{X}_{i}} \rangle \langle \psi_{\Delta \tilde{X}_{i}} | \} \notag \\
	&= \frac{1}{\sum_{j=1}^{k} \sum_{m=1}^{D} | x_{mi} - x^{(i)}_{mj} |^{2} } \sum_{j, j^{'}=1}^{k} \sum_{m=1}^{D} (x_{mi} - x^{(i)}_{mj}) (x_{mi} - x^{(i)}_{mj^{'}})^{\ast} | j \rangle \langle j^{'} | \notag \\
	&= \frac{\tilde{C}_{i}}{\mathrm{tr} \tilde{C}_{i}},
\end{align}
where $\tilde{C}_{i} = \Delta \tilde{X}^{T}_{i} \Delta \tilde{X}_{i}$.

In the last step, we attempt to construct the weight matrix $W$ by repeatedly solving the linear equation $C_{i}W_{i} = \textbf{1}_{N}$ for $i = 1, 2, \dots, N$. Before applying the quantum algorithm for linear systems of equations (HHL)~\cite{7}, we should firstly extend the matrix $\tilde{C}_{i}$ to $C_{i} = \begin{bmatrix}
	\tilde{C}_{i} & 0 \\
	0 & 0 
\end{bmatrix}_{N \times N}$ which is obviously $k$-sparse. Finally, as a matter of fact, our goal is to solve $| W_{i} \rangle = C_{i}^{-1} | \textbf{1}_{N} \rangle$ with $| W_{i} \rangle = (W_{i1}, W_{i2}, \dots, W_{ik}, 0, \dots, 0)^{T}_{N}$ and $| \textbf{1}_{N} \rangle = \frac{1}{\sqrt{N}}(1, 1, \dots, 1)_{N}^{T}$. We input the state $| \psi_{init} \rangle = | 0 \rangle^{R_{1}} | \underbrace{0, \dots, 0}_{n} \rangle ^{C_{1}} | \textbf{1}_{N} \rangle^{B_{1}}$ where $n = \log N$, and perform the HHL algorithm on it to obtain $| W_{i} \rangle$. Before detailedly describing the procedure of solving $| W_{i} \rangle$, we generalize the HHL algorithm at first.
 
\begin{figure}
\centering
\includegraphics[width=\textwidth]{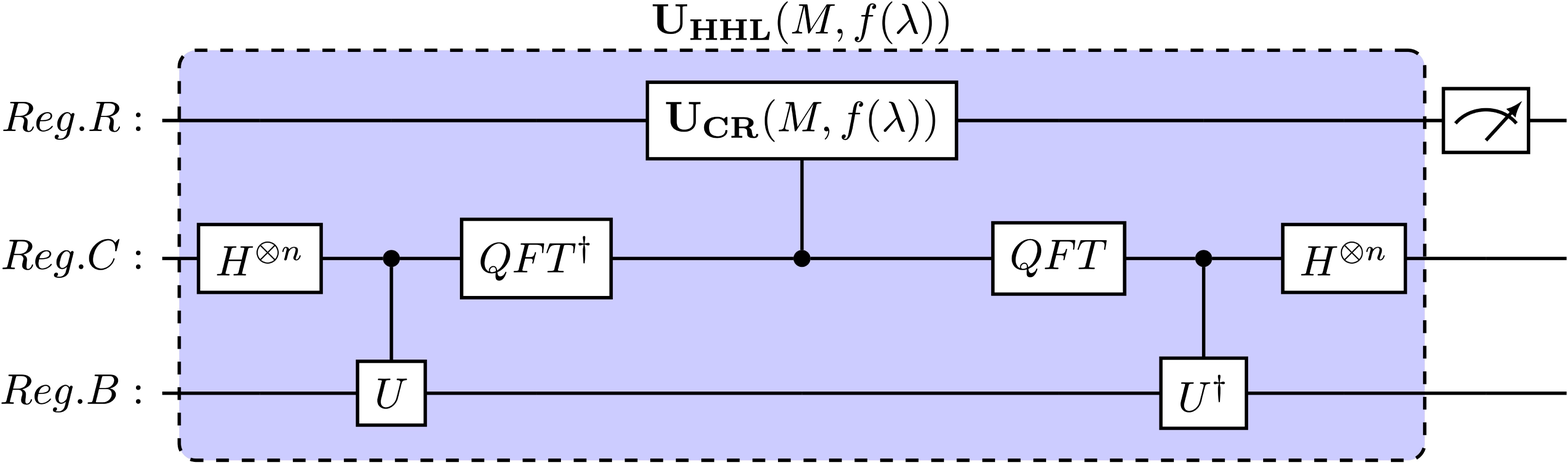}
\caption{The quantum circuit of the HHL algorithm \cite{28}}
\label{fig:U_HHL}
\end{figure}

\begin{figure}
\centering
\includegraphics[width=0.6\textwidth]{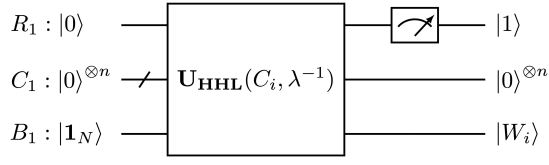}
\caption{The quantum circuit of computing $W_{i}$}
\label{fig:W}
\end{figure}

As shown in Fig. \ref{fig:U_HHL} \cite{28}, $\textbf{U}_{\textbf{HHL}}(M, f(\lambda)) = (\textbf{I}^{R} \otimes \textbf{U}^{\dagger}_{\textbf{PE}}(M))(\textbf{U}_{\textbf{CR}}^{RC}(M, f(\lambda)) \otimes \textbf{I}^{B})(\textbf{I}^{R} \otimes \textbf{U}_{\textbf{PE}}(M))$ represents the unitary evolution of the HHL algorithm where the unitary operation $\textbf{U}_{\textbf{PE}}(M) = (\textbf{QFT}^{\dagger} \otimes \textbf{I}^{B}) \left( \sum_{\tau=0}^{T-1} | \tau \rangle \langle \tau |^{C} \otimes e^{i M \tau t_{0} / T} \right)(\textbf{H}^{\otimes t} \otimes \textbf{I}^{B})$ represents the phase estimation algorithm performed~\cite{29} with a specified sparse matrix $M$, $\textbf{QFT}^{\dagger}$ stands for the inverse QFT, and $\textbf{U}_{\textbf{CR}}(M, f(\lambda))$ represents a conditional rotation operation which is specifically
\begin{equation}\label{eq:U_{CR}}
	| 0 \rangle^{R} \otimes | \lambda \rangle^{C} \rightarrow \left( \sqrt{1 - \gamma^{2} f(\lambda)^{2}} | 0 \rangle + \gamma f(\lambda) | 1 \rangle \right)^{R} \otimes | \lambda \rangle^{C},
\end{equation}
where the register $R$ is controlled by the register $C$; $\gamma$ is a constant; $\lambda$ represents the eigenvalue of $M$ and $f(\lambda)$ is a specified function about $\lambda$. Thus, the overall procedure of solving $| W_{i} \rangle$ is presented as follows:
\begin{align}\label{W_{i}}
	| \psi_{init} \rangle &\xrightarrow{\textbf{U}_{\textbf{PE}}(C_{i})} |0 \rangle^{R_{1}} \sum_{i=1}^{N} \langle u_{1i} | \textbf{1}_{N} \rangle | \lambda_{1i} \rangle^{C_{1}} | u_{1i} \rangle^{B_{1}} \notag \\
	&\xrightarrow{\textbf{U}_{\textbf{CR}}(C_{i}, \lambda^{-1})} (\sqrt{1 - \frac{\gamma_{1}^{2}}{\lambda_{1i}^{2}}} | 0 \rangle + \frac{\gamma_{1}}{\lambda_{1i}} | 1 \rangle)^{R_{1}} \sum_{i=1}^{N} \langle u_{1i} | \textbf{1}_{N} \rangle | \lambda_{1i} \rangle^{C_{1}} | u_{1i} \rangle^{B_{1}} \notag \\ 
	&\xrightarrow{\textbf{Uncompute}} (\sqrt{1 - \frac{\gamma_{1}^{2}}{\lambda_{1i}^{2}}} | 0 \rangle + \frac{\gamma_{1}}{\lambda_{1i}} | 1 \rangle)^{R_{1}} \sum_{i=1}^{N} \langle u_{1i} | \textbf{1}_{N} \rangle | u_{1i} \rangle^{B_{1}} \notag \\
	&\xrightarrow{\textbf{Measurement}} | W_{i} \rangle = \frac{1}{\sqrt{\sum_{i=1}^{N}\gamma_{1}^{2} / \lambda_{1i}^{2}}} \sum_{i=1}^{N} \frac{\gamma_{1}}{\lambda_{1i}} \langle u_{1i} | \textbf{1}_{N} \rangle | u_{1i} \rangle,
\end{align}
where $\gamma_{1}$ is a constant; $\lambda_{1i}$ are the eigenvalues of $C_{i}$ and $u_{1i}$ are the corresponding eigenvectors. By repeating this process for $i = 1, 2, \dots N$, we can acquire all the columns of $W$. The quantum circuit of solving $W_{i}$ is depicted in Fig.~\ref{fig:W}.

\subsection{Preparation of the target matrix $M$}
\label{subsec:preparation_of_M}
\begin{figure}
\centering
\includegraphics[width=0.6\textwidth]{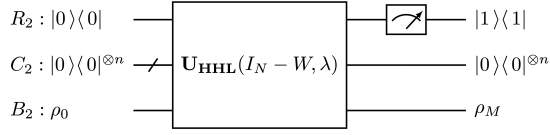}
\caption{The quantum circuit of computing the target matrix $M$}
\label{fig:M}
\end{figure}

Almost the same as the preparation of $| x_{i} - x_{j}^{(i)} \rangle$, we input the quantum states $| e_{i} \rangle$ and $| W_{i} \rangle$ into the quantum subtractor circuit resulting in $| e_{i} - W_{i} \rangle$ where $e_{i}$ is the $i$th column of the identity matrix $I_{N}$. We input the state $| 0 \rangle | e_{i} \rangle | W_{i} \rangle$ into the swap test circuit and trace out the $| W_{i} \rangle$ register. The probability of achieving $| 0 \rangle$ after measuring the ancilla is $P_{2}(0) = \frac{1}{2} + \frac{1}{2} \textrm{Re}\langle e_{i} | W_{i} \rangle$. Thus, the norm 
\begin{align}\label{eq:norm_e_i-w_i}
	| e_{i} - W_{i} | &= \sqrt{2 - 2 \textrm{Re} \langle e_{i} | W_{i} \rangle} \notag \\
	&= 2 \sqrt{1 - P_{2}(0)}.
\end{align}
Having acquired $| e_{i} - W_{i} \rangle$ and $| e_{i} - W_{i} |$, we have quantum access to the matrix $I_{N} - W = \sum_{i = 1}^{N} | e_{i} - W_{i} | | e_{i} - W_{i} \rangle \langle i |$. In the next, we apply the $\textbf{U}_{\textbf{HHL}}(I_{N} - W, \lambda)$ on the quantum state $\rho_{0} = \frac{1}{\sqrt{N}} \sum_{i=1}^{N} | i \rangle \langle i |$ to obtain the density operator $\rho_{M}$ which is proportional to the target matrix $M = (I_{N} - W)(I_{N} - W)^{T}$ according to section~\ref{sec:classical_LLE}. Specifically, this procedure~\cite{19} can be summarized as follows:
\begin{align}\label{eq:HHL_rho_M}
	&|0 \rangle \langle 0 |^{R_{2}} \otimes {| 0 \rangle \langle 0 |^{\otimes n}}^{C_{2}} \otimes \rho_{0}^{B_{2}} \notag \\
	&\xrightarrow{\textbf{U}_{\textbf{PE}}(I_{N}-W)} \sum_{i, j=1}^{N} \langle u_{2i} | \rho_{0} | u_{2j} \rangle | 0 \rangle \langle 0 |^{R_{2}} \otimes | \lambda_{2i} \rangle \langle \lambda_{2j} |^{C_{2}} \otimes | u_{2i} \rangle \langle u_{2j} |^{B_{2}} \notag \\
	&\xrightarrow{\textbf{U}_{\textbf{CR}}(I_{N} - W, \lambda)} \sum_{i, j=1}^{N} \langle u_{2i} | \rho_{0} | u_{2j} \rangle | \psi_{anc} \rangle \langle \psi_{anc^{'}} |^{R_{2}} \otimes | \lambda_{2i} \rangle \langle \lambda_{2j} |^{C_{2}} \otimes | u_{2i} \rangle \langle u_{2j} |^{B_{2}} \notag \\
	&\xrightarrow{\textbf{Uncompute}} \sum_{i, j=1}^{N} \langle u_{2i} | \rho_{0} | u_{2j} \rangle | \psi_{anc} \rangle \langle \psi_{anc^{'}} |^{R_{2}} \otimes | u_{2i} \rangle \langle u_{2j} |^{B_{2}} \notag \\
	&\xrightarrow{\textbf{Measurement}} \rho_{M} = \frac{1}{\sqrt{\sum_{i, j=1}^{N} \gamma_{2}^{2} \gamma_{2^{'}}^{2} \lambda_{2i}^{2} {\lambda_{2j}^{\ast}}^{2}}} \sum_{i, j=1}^{N} \gamma_{2} \gamma_{2^{'}} \lambda_{2i} \lambda_{2j}^{\ast} \langle u_{2i} | \rho_{0} | u_{2j} \rangle |u_{2i} \rangle \langle u_{2j}|,
\end{align} 
where the ancilla
\begin{equation}\label{eq:psi_1}
	\begin{cases}
	| \psi_{anc} \rangle = \sqrt{1 - \gamma_{2}^{2} \lambda_{2i}^{2}} | 0 \rangle + \gamma_{2} \lambda_{2i} | 1 \rangle; \\
	\langle \psi_{anc^{'}} | = \sqrt{1 - \gamma_{2^{'}}^{2} {\lambda^{\ast}_{2j}}^{2}} \langle 0 | + \gamma_{2^{'}} \lambda_{2j}^{\ast} \langle 1 |;
	\end{cases}
\end{equation}
$\gamma_{2}$, $\gamma_{2^{'}}$ are constants; $\lambda_{2i}$ are the eigenvalues of $I_{n} - W$ and $u_{2i}$ are the corresponding eigenvectors. In addition, the quantum circuit of this procedure is presented in~Fig. \ref{fig:M}.

It is worth mentioning that we can also compute the $\rho_{M}$ utilizing the qRAM just like the preparation of $\tilde{C}_{i}$. We can firstly construct the state 
\begin{equation}\label{eq:psi_{I_{n}-W}}
	| \psi_{I_{N} - W} \rangle = \frac{1}{\sqrt{\sum_{i=1}^{N}\sum_{m=1}^{N} | e_{mi} - W_{mi} |^{2}}} \sum_{i=1}^{N} \sum_{m=1}^{N} (e_{mi} - W_{mi}) | i \rangle | m \rangle, 
\end{equation} 
and trace out the $| i \rangle$ register which is different from the Eq.~\eqref{eq:rho_{tilde_C_{i}}} resulting in 
\begin{align}\label{eq:partial_trace_rho_{M}}
	\rho_{M} &= \mathrm{tr}_{i} \{ | \psi_{I_{N} - W} \rangle \langle \psi_{I_{N} - W} | \} \notag \\
	&= \frac{1}{\sum_{i=1}^{N} \sum_{m=1}^{N} | e_{mi} - W_{mi} |^{2}} \sum_{i=1}^{N} \sum_{m, m^{'}=1}^{N} (e_{mi} - W_{mi})(e_{mi} - W_{mi})^{\ast} | m \rangle \langle m^{'} | \notag \\
	&= \frac{M}{\mathrm{tr}M}.
\end{align}

Ultimately, we obtain the density operator $\rho_{M}$ according to Eq.~\eqref{eq:HHL_rho_M} or Eq.~\eqref{eq:partial_trace_rho_{M}}. Our goal is to solve $M$'s second to $(d+1)$th smallest eigenvectors which are actually exactly the same as $\rho_{M}$'s corresponding eigenvectors. Thus, $\rho_{M}$ has completely met our subsequent needs.

\subsection{Computation of the embedding coordinates $Y$}
\label{subsec:compute_Y}
Before applying the quantum linear algebra techniques, some modifications should be made. We define the matrix $J = \xi I_{N} - \rho_{M}$ for some large constant $\xi$. According to the analysis in section~\ref{subsec:preparation_of_M}, it is obvious that solving the eigenvectors which are corresponding to $M$'s second to $(d+1)$th smallest eigenvalues is equivalent to finding the eigenvectors which are corresponding to $J$'s second to $(d+1)$th largest eigenvalues. Hence, our goal at present is to reveal the eigenvectors corresponding to the second to $(d+1)$th dominant eigenvalues of $J$. 

Because $J = \xi I_{N} - \rho_{M}$, the exponentiation of $J$
\begin{equation}\label{eq:exponentiate_J}
	e^{-i J \Delta t} = e^{-i \xi I_{N} \Delta t} e^{i \rho_{M} \Delta t} + O({\Delta t}^{2}).
\end{equation}
The matrix $\xi I_{N}$ is trivial. The Hamiltonian simulation can be efficiently implemented by the trick in Ref.~\cite{18} as follows
\begin{align}
	\mathrm{tr}_{1} \{ e^{-iS\Delta t} \rho_{M} \otimes \sigma e^{iS\Delta t}\} &= \sigma - i\Delta t[\rho_{M},\sigma] + O({\Delta t}^2) \notag \\
	&\approx e^{-i \rho_{M} \Delta t} \sigma e^{i \rho_{M} \Delta t}
	\label{eq:qpca}
\end{align}
where $S = \sum_{i, j=1}^{N} | i \rangle \langle j | \otimes | j \rangle \langle i |$ is the swap operator; $\sigma$ is a specified density operator; $\mathrm{tr}_{1}$ is the partial trace over the first register; $\Delta t = t / L$ for some large $L$.

Given $L$ copies of $\rho_{M}$, the controlled $e^{-i S \Delta t}$ operation is $CU_{S} = | 0 \rangle \langle 0 | \otimes I + | 1 \rangle \langle 1 | \otimes e^{-i S \Delta t}$. Apply the controlled unitary operation
\begin{equation}\label{eq:exponentiate_rho_M}
	CU_{M} = \sum_{l = 0}^{L} | l \Delta t \rangle \langle l \Delta t | \otimes e^{i l \rho_{M} \Delta t} \sigma e^{-i l \rho_{M} \Delta t} 
\end{equation}
on $\sigma \otimes \rho_{M}^{\otimes L}$ where $\sigma = | u \rangle \langle u |$ corresponding to $\rho_{M}$'s eigenvectors.

With the input state $\rho_{M} = \sum_{i} \lambda_{i} | u_{i} \rangle \langle u_{i} |$ where $u_{i}$ are the eigenvectors of $\rho_{M}$ and $\lambda_{i}$ are the corresponding eigenvalues, the whole procedure above is 
\begin{equation}
	(CU_{M})^{L} | L \Delta t \rangle | u_{M} \rangle \rightarrow | L \Delta t \rangle e^{i\rho_{M} t} | u_{M} \rangle.
	\label{eq:Hamiltonian simulation}
\end{equation}
We can embed this controlled Hamiltonian simulation process into the phase estimation algorithm $\textbf{U}_{\textbf{PE}}(J)$ resulting in $\sum_{i=1}^{N} \lambda_{i}^{(J)} | u_{i}^{(J)} \rangle \langle u_{i}^{(J)} | \otimes | \lambda_{i}^{(J)} \rangle \langle \lambda_{i}^{(J)} |$. The second to $(d+1)$th eigenvectors corresponding the dominant eigenvalues of $J$ can be obtained through sampling from this state. Subsequently, the $2$nd to $(d + 1)$th smallest eigenvectors of $\rho_{M}$ can be obtained. Ultimately, the final $d$-dimensional data set $Y = \{ y_{i} \in \mathbb{R}^{d}: 1 \leq i \leq N \} = (u_{2}, \dots, u_{d+1})^{T}$.

\begin{algorithm}[t]
	\caption{Linear-algebra-based QLLE}
	\KwIn{High-dimensional data $X$.}
	\KwOut{Low-dimensional data $Y$.}
	\emph{step 1}: Preprocess $X$ by the quantum $k$-NN algorithm to obtain the $k$ nearest neighbors of each data point of $X$. \\
	\emph{step 2}: Compute the weight matrix $W$ as in section~\ref{subsec:construction_of_W}. \\
	\emph{step 3}: Prepare the target matrix $M$ as in section~\ref{subsec:preparation_of_M}. \\
	\emph{step 4}: Compute the low-dimensional data $Y$ as in section~\ref{subsec:compute_Y}.	
	\label{alg:LABQLLE}
\end{algorithm}

\subsection{Algorithmic Complexity analysis}
\label{subsec:complexity}
To evaluate the performance of the linear-algebra-based QLLE, the algorithmic complexity analysis of the classical LLE and the linear-algebra-based QLLE are described in this subsection.

The classical LLE algorithm mainly contains three steps. Firstly, we want to find the $k$ nearest neighbors of each input data point. Generally, the classical $k$-NN algorithm needs to construct the corresponding data structure index in $O(N \log N)$ and then to search the $k$ nearest neighbors in $O(k \log N)$~\cite{30}. Then, the classical LLE subsequently attempts to solve $N$ set of linear equations in size of $k \times k$ requiring computational complexity in $O(N k^{3})$. Finally, the corresponding complexity of the eigenvalue analysis is in $O(d N^2)$. In summary, the overall complexity of the classical LLE algorithm is the sum of all these operations mentioned above~\cite{31}.

In contrast with the classical LLE algorithm, we can obtain the computational complexity of the linear-algebra-based QLLE algorithm as follows. In data preprocessing, we invoke the quantum $k$-NN algorithm~\cite{21} to find out the $k$ nearest neighbors of the original data in $O(R \sqrt{k N})$ where $R$ represents the Oracle execution times. Compared with the classical $k$-NN algorithm, the quantum $k$-NN algorithm can achieve quadratic speedup. As to the main part of the algorithm, we solve the weight matrix $W$ by performing the HHL~\cite{7} $N$ times on $C_{i}$ for $i = 1, 2, \dots, N$ in $O(\sum_{i=1}^{N} \kappa_{i}^{2} \log N / \epsilon_{1})$ where $\epsilon_{1}$ is the error parameter and $\kappa_{i}$ is the condition number of $C_{i}$. Subsequently, the target matrix $M$ can be prepared in $O(\kappa_{0}^{4} \log N / \epsilon_{2}^{3})$ where $\kappa_{0}$ is the condition number of the matrix $I_{N}-M$ and $\epsilon_{2}$ is the tolerance error~\cite{19}. It is worth mentioning that this quantum matrix multiplication operation for preparing $M$ achieves exponential speedup compared with the classical matrix multiplication in $O(N^{3})$. In the last step, the $d$ principal components can be obtained with performing the qPCA algorithm~\cite{18} on the target matrix $M$ in $ O(d \log D)$.

Therefore, in addition to the data preprocessing where the quantum $k$-NN algorithm can quadratically reduce the resources in contrast to the classical $k$-NN algorithm, the linear-algebra-based QLLE algorithm can achieve exponential speedup compared with the classical LLE algorithm in each steps resulting in the overall exponential acceleration.

\section{Variational quantum locally linear embedding}
\label{sec:VQLLE}
In addition to the linear-algebra-based QLLE, we can alternatively implement the QLLE on the near term quantum devices with a variational hybrid quantum-classical procedure. Firstly, the weight matrix is constructed by a variational quantum linear solver~\cite{32}. Subsequently, two different configurations of computing the low-dimensional data $Y$ are presented. The concrete steps of the VQLLE are presented as follows. 

\subsection{Computation of the weight matrix $W$}
\label{subsec:compute_W}
In this subsection, the weight matrix $W$ is computed through a variational hybrid quantum-classical procedure. Given the original high-dimensional data matrix $X = (x_{1}, x_{2}, \cdots, x_{N}) \in \mathbb{R}^{D \times N}$, the $k$ nearest neighbors of each data sample $x_{i}$ can be obtained by the $k$-NN algorithm. The local covariance matrix $C_{i}$ can be achieved with $x_{i}$ and the corresponding $k$ nearest neighbors $\{ x_{j}^{(i)} \}$ for $j = 1, 2, \cdots, k$. Subsequently, we design the ansatz states $| \psi(\theta_{w_{i}}) \rangle$ with a set of parameters $\{ \theta_{w_{i}} \}$ for $i = 1, 2, \cdots, N$. As introduced in section~\ref{sec:classical_LLE}, the $i$th column $W_{i}$ of the weight matrix $W$ can be computed by solving the linear system of equation as Eq.~\eqref{eq:W_{i}}. Essentially, it is equivalent to prepare the quantum state 
\begin{equation}
	| \psi_{vi} \rangle = \frac{C_{i} | \psi(\theta_{wi}) \rangle}{\sqrt{\langle \psi(\theta_{w_{i}}) | C_{i}^{\dagger} C_{i} | \psi(\theta_{w_{i}}) \rangle}} 
	\label{eq:psi_vi}
\end{equation}
to be proportional to $| \textbf{1}_{N} \rangle$. By minimizing the cost function
\begin{equation}
	L_{1}(\theta_{w}) = \frac{1}{N}\sum_{i=1}^{N} \left \vert \frac{\langle \textbf{1}_{N} | C_{i} | \psi(\theta_{w_{i}})  \rangle}{\sqrt{\langle \psi(\theta_{w_{i}}) | C_{i}^{\dagger} C_{i} | \psi(\theta_{w_{i}}) \rangle}} \right \vert^{2}
	\label{eq:L_1}
\end{equation}
the optimal ansatz states $| \psi(\theta_{w_{i}}^{\ast}) \rangle$ representing the $i$th column of the weight matrix $W_{i}$ can be obtained. Hence, the weight matrix $W = \sum_{i=1}^{N}| \psi(\theta_{w_{i}}^{\ast}) \rangle \langle i |$ can be achieved.

\subsection{Computation of the low-dimensional data $Y$}
\label{subsec:computation_of_Y}
In this subsection, the computation of the low-dimensional data $Y$ can be implemented in two different configurations. 

For the first design, the computation is implemented in the end-to-end way. The end-to-end implementation means that only the input and the output are cared with the intermediate process ignored. The given input is processed to the target output with a parameterized quantum circuit. As introduced in section~\ref{sec:classical_LLE}, we define the cost function 
\begin{equation}
	L_{2}(\theta_{y}) = \vert | Y^{T} \rangle - W | Y^{T} \rangle \vert^{2}, 
	\label{eq:L_2}
\end{equation} 
where $| Y^{T} \rangle = \sum_{i=1}^{N} | y_{i}(\theta_{y}) \rangle | i \rangle$ is made up of the ansatz states $| y_{i}(\theta_{y}) \rangle$ with a set of parameters $\{ \theta_{y} \}$. By minimizing the cost function $L_{2}$ with a classical optimization algorithm, the optimal ansatz states $| Y^{T}_{\ast} \rangle = \sum_{i=1}^{N} | y_{i}(\theta_{y}^{\ast}) \rangle | i \rangle$ can be obtained. Therefore, the low-dimensional data $Y = \sum_{i=1}^{N} | y_{i}(\theta_{y}^{\ast}) \rangle \langle i |$.  

For the second design, the computation of $Y$ is based on the variational quantum eigensolver (VQE) inspired from Ref.~\cite{15,16}. According to subsection~\ref{subsec:compute_W}, the quantum state representing the matrix $I_{N} - W$ is $| \psi_{I_{N} - W} \rangle = \sum_{i=1}^{N} | i \rangle | e_{i} - \psi_{\theta_{w_{i}}^{\ast}} \rangle$ and $\rho_{M} = \mathrm{tr}_{i} \{ | \psi_{I_{N} - W} \rangle \langle \psi_{I_{N} - W} | \}$. The specific steps of computing $Y$ in this configuration are presented as follows.

(1) Prepare the ansatz states $| \psi(\lambda_{\theta_{e}}) \rangle$. As a matter of fact, the ansatz states $| \psi(\lambda_{\theta_{e}}) \rangle$ represents a set of parameterized circuits in practice where $\{\theta_{e}\}$ represents a set of parameters. The ansatz states are prepared to introduce parameters to the cost function $E(\lambda_{\theta_{e}})$. Specifically, we apply some parameterized quantum rotation operations on the ground states and entangle all the input states together resulting in the final ansatz states. The specific quantum circuit of preparing the ansatz states is presented as in Ref.~\cite{33}.

(2) Construct the cost function $L_{3}(\lambda_{\theta_{e}})$. The cost function 
\begin{equation}\label{eq:L_3}
	L_{3}(\lambda_{\theta_{e}}) = \langle \psi(\lambda_{\theta_{e}}) | \rho_{M} | \psi(\lambda_{\theta_{e}}) \rangle + \sum_{i=1}^{d+1} \alpha_{i}|\langle \psi(\lambda_{\theta_{e}}) | \psi(\lambda_{i}) \rangle |^{2},
\end{equation}
where $\alpha_{i}$ are the corresponding coefficients~\cite{16}. The first term of Eq.~\eqref{eq:L_3} called the expectation value term $\langle \psi(\lambda_{\theta_{e}}) | \rho_{M} | \psi(\lambda_{\theta_{e}}) \rangle$ can be estimated with a one-step phase estimation circuit~\cite{34}. The expectation value is achieved with a low-depth quantum circuit in $O(\epsilon^{\mu}) (\mu \geq 0)$ and measurements in $O(1 / \epsilon^{2 + \nu}) (0 < \nu \leq 1)$ to precision $\epsilon$. Essentially, the computation of the overlap term is to estimate the inner product of $| \psi(\lambda_{\theta_{e}}) \rangle$ and $| \psi(\lambda_{i}) \rangle$. Thus, the overlap term $|\langle \psi(\lambda_{\theta_{e}}) | \psi(\lambda_{i}) \rangle |^{2}$ can be estimated with the swap test circuit~\cite{26} utilized to compute the inner product of two quantum states in $O(1 / \epsilon^{2})$ measurements and $O(\log{N})$ circuit depth. Ref.~\cite{35} proposed a variational parameterized quantum circuit to compute the inner product demonstrating that the inner product of two complicate kernel states can also be computed efficiently. In addition, the destructive swap test~\cite{30} which exhibits promotion performance compared with the original swap test can compute the overlap term with an $O(1)$-depth quantum circuit.  

(3) Invoke classical operations. Having finished all the quantum parts, we invoke the classical adder to sum over all the expectation value terms and the overlap terms resulting in the cost function $L_{3}(\lambda_{\theta_{e}})$. Subsequently, $L_{3}(\lambda_{\theta_{e}})$ can be minimized with a classical optimizer to obtain the eigenvalue $\lambda_{\theta_{e}}$.

(4) Iterate the steps (1) to (3), we can obtain the second to $d+1$th smallest eigenvalues of $\rho_{M}$ and their corresponding eigenvectors. 

Therefore, the locally linear embedding can be implemented by the VQLLE with a variational hybrid quantum-classical procedure. The VQLLE combines quantum circuits with the classical optimization algorithm to implement the procedure of LLE on the near term quantum devices.

\begin{algorithm}[t]
	\caption{Variational quantum locally linear embedding}
	\KwIn{High-dimensional data $X$.}
	\KwOut{Low-dimensional data $Y$.}
	\emph{step 1}: Preprocess $X$ by the $k$-NN algorithm to obtain the corresponding $k$ nearest neighbors of each data sample of $X$ and compute the local covariance $C_{i}$. \\
	\emph{step 2}: Minimize the cost function $L_{1}(\theta_{w})$ as Eq.~\eqref{eq:L_1} to obtain the weight matrix $W$. \\
	\emph{step 3}: Minimize the cost function $L_{2}(\theta_{y})$ as Eq.~\eqref{eq:L_2} to obtain the low-dimensional data $Y$. \\
	\emph{or} \\
	\emph{step 4}: Minimize $L_{3}(\lambda_{\theta_{e}})$ as Eq.~\eqref{eq:L_3} to obtain the $2$nd to $(d + 1)$th smallest eigenvalues and the corresponding eigenvectors of $\rho_{M}$ to construct the low-dimensional data $Y$. 
	\label{alg:VQLLE}
\end{algorithm}

\section{Numerical experiments}
\label{sec:numerical experiments}
In this section, the numerical experiments of the algorithms proposed in our work will be presented. In our simulation, two representative data sets in the field of dimensionality reduction, the Swiss roll data set and the S-curve data set, are selected. The experiments in this paper are implemented on a classical computer with the Python programming language, the Scikit-learn machine learning library~\cite{36} and the Qiskit quantum computing framework~\cite{37}. The numerical experiments present the simulation results on applying the linear-algebra-based QLLE and the VQLLE to the two data sets respectively.

\begin{figure*}[tb]
	\centering
	\subfigure[]{
    \includegraphics[width=0.3\textwidth]{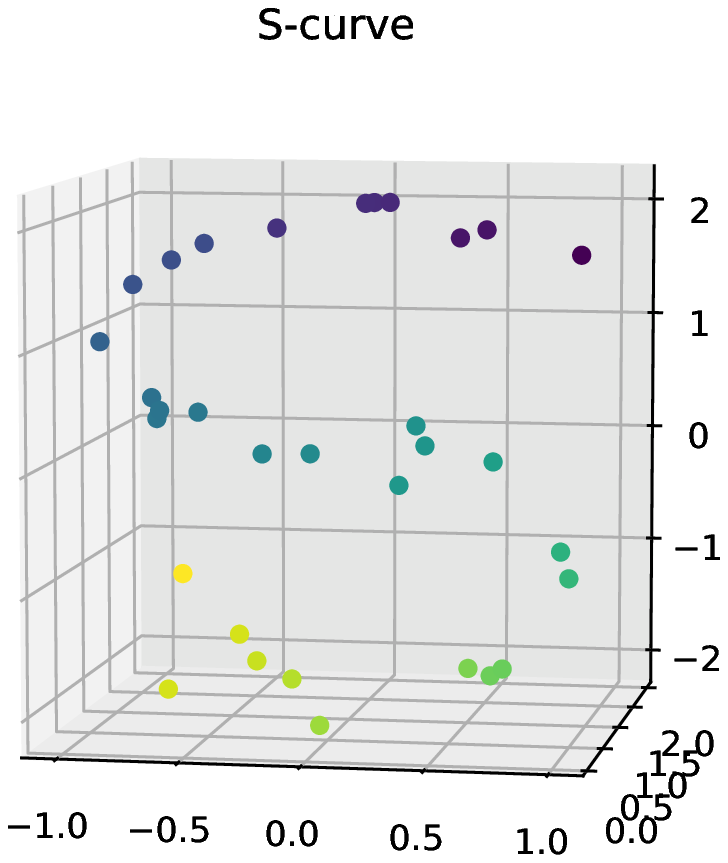}
	\label{fig:7_1}
	}
	\quad
	\subfigure[]{
	\includegraphics[width=0.3\textwidth]{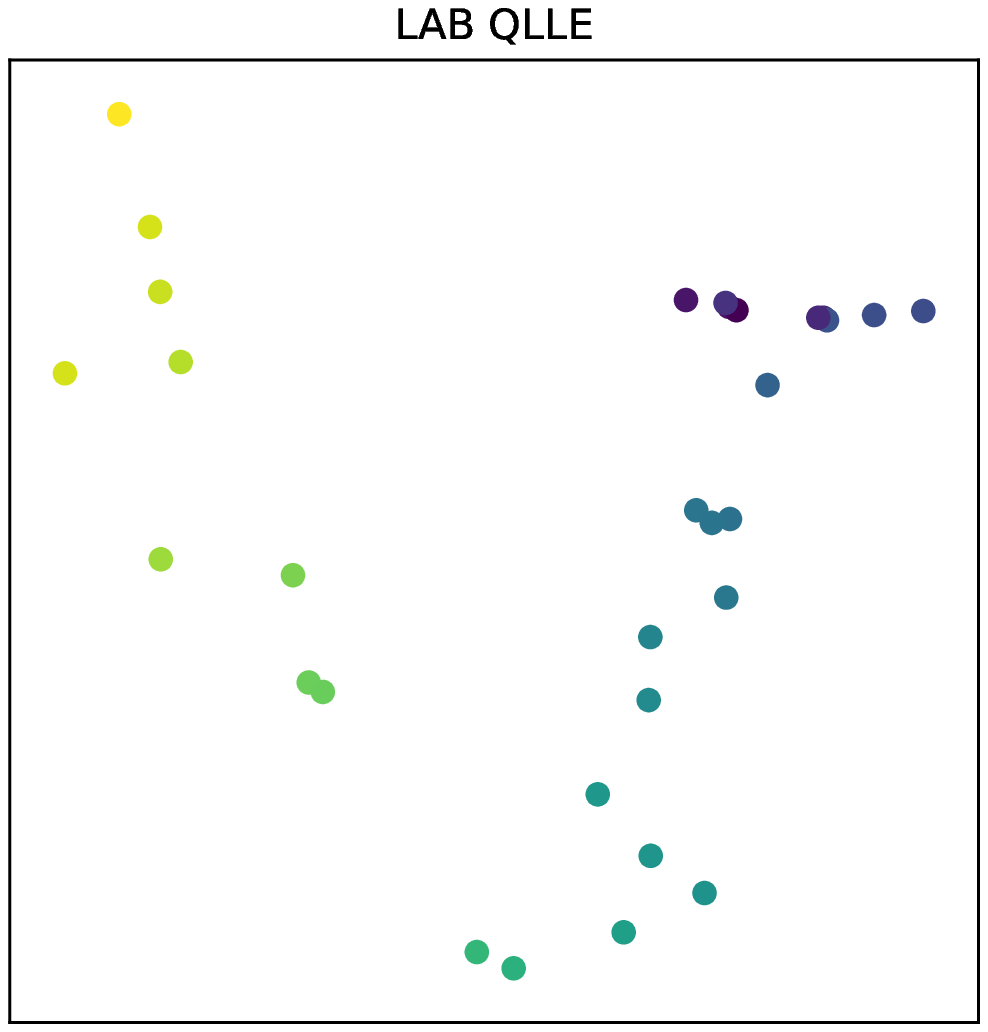}
	\label{fig:7_2}
	}
	\quad
	\subfigure[]{
	\includegraphics[width=0.3\textwidth]{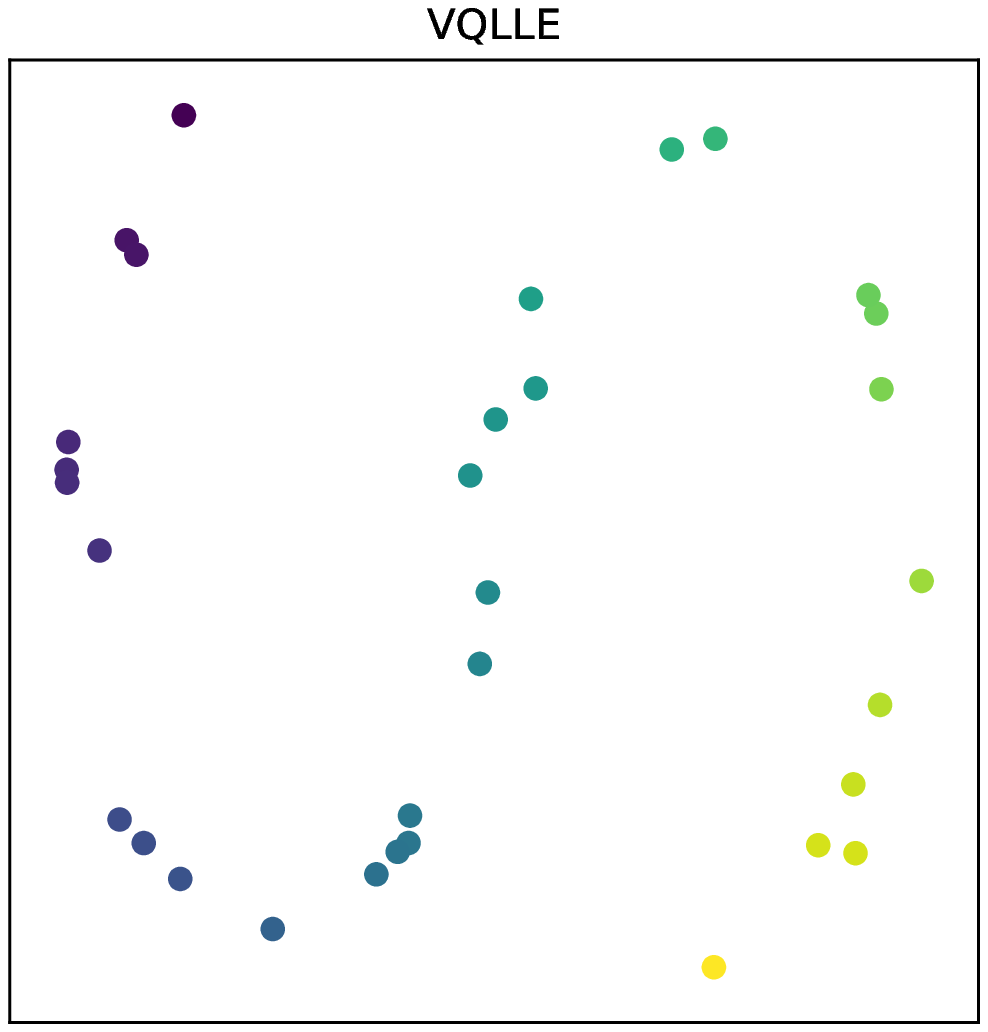}
	\label{fig:7_3}
	}
	
	\subfigure[]{
	\includegraphics[width=0.3\textwidth]{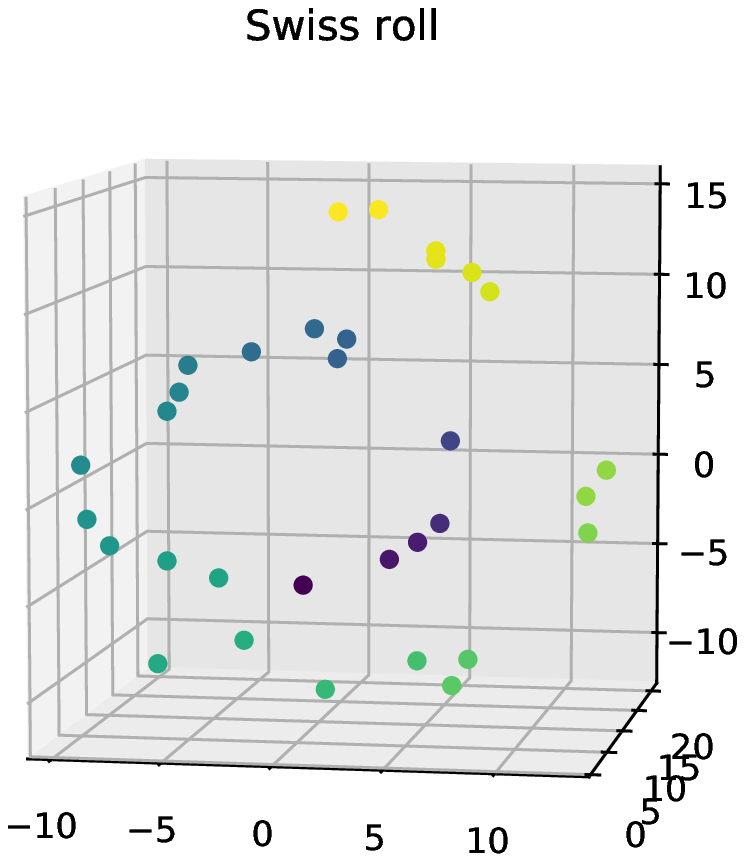}
	\label{fig:7_4}
	}
	\quad
	\subfigure[]{
	\includegraphics[width=0.3\textwidth]{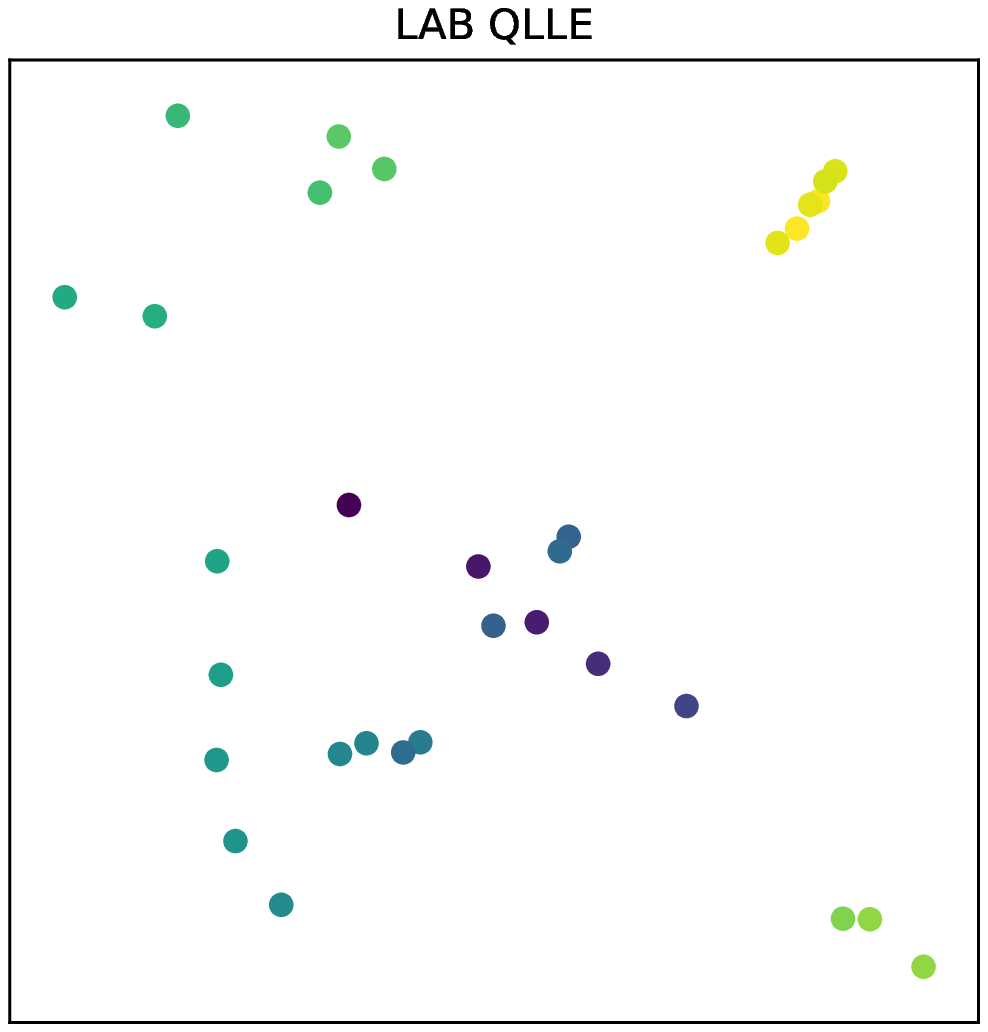}
	\label{fig:7_5}
	}
	\quad
	\subfigure[]{
	\includegraphics[width=0.3\textwidth]{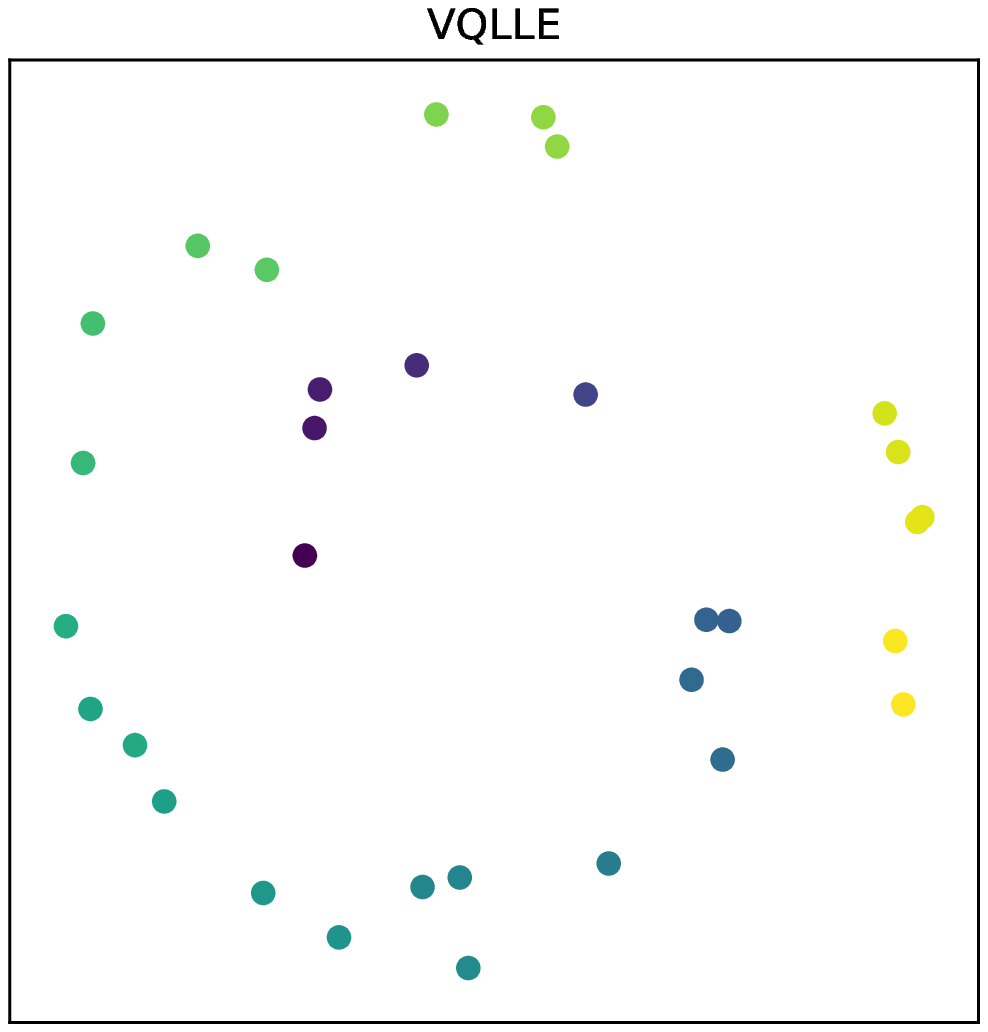}
	\label{fig:7_6}
	}
	
	\caption{The simulation results of the linear-algebra-based QLLE and the VQLLE applied on the S-curve data set and the Swiss roll data set respectively. (a) The S-curve data set. (b) The result of applying the linear-algebra-based QLLE on the S-curve. (c) The result of applying the VQLLE on the S-curve. (d) The Swiss roll data set. (e) The result of applying the linear-algebra-based QLLE on the Swiss roll. (f) The result of applying the VQLLE on the Swiss roll}
	\label{fig:7}
\end{figure*}

\subsection{Data sets}
\label{subsec:data sets}
In our experiments, the Swiss roll data set and the S-curve data set are selected as the benchmark data sets. The data samples of the two data sets are distributed on manifold surfaces of different shapes as shown in Fig.~\ref{fig:7_1}, Fig.~\ref{fig:7_4} respectively. In order to demonstrate the effect of the dimensionality reduction while ensuring the feasibility of the QLLE algorithms, we sample $32$ three-dimensional data points from each manifold as the original high-dimensional data sets. 

\subsection{Experiments on the linear-algebra-based QLLE}
\label{subsec:simulation of the QLLE}
Although the linear-algebra-based QLLE can achieve quadratic speedup in the number and dimension of the given data compared with the classical LLE, it actually requires high-depth quantum circuits and fully coherent evolution. In our experiments, the number of the nearest neighbors $k = 4$. However, the required circuit depth is over $100$ which is relatively large. As shown in Fig.~\ref{fig:7_2} and Fig.~\ref{fig:7_5}, the linear-algebra-based QLLE projects the given three-dimensional S-curve and Swiss roll data to the two-dimensional space respectively. The results demonstrate that the linear-algebra-based QLLE can achieve the dimensionality reduction with preserving the local topological structure of the original data. However, the global data characteristics are not well reflected. In addition, the depth grows rapidly and the fidelity deteriorates dramatically as the $k$ increases. Thus, the linear-algebra-based QLLE is proved to have the potential to achieve the dimensionality reduction with quantum speedup over the corresponding classical algorithm, but it is not achievable to perform the linear-algebra-based QLLE on the noisy intermediate-scale quantum devices.

\subsection{Experiments on the VQLLE}
\label{subsec:VQLLE}
The implementation of the VQLLE proposed in section~\ref{sec:VQLLE} mainly contains two parts, the quantum circuits and the classical optimization. For the quantum part, the parameterized quantum circuits as presented in Ref.~\cite{33} are designed to construct the cost function. Subsequently, the classical optimization algorithm, the AdaGrad~\cite{38}, is invoked to minimize the cost function to obtain the optimal parameters resulting in the target low-dimensional data $Y$. The results of the VQLLE presented in Fig.~\ref{fig:7_3} and Fig.~\ref{fig:7_6} demonstrates that the VQLLE can also achieve the procedure of the dimensionality reduction with preserving both the local and the global topological structures of the given data. Although the VQLLE can not be implemented with quantum speedup in the whole procedure, it efficiently combines the quantum and classical computation to realize the dimensionality reduction on the near term quantum devices. In addition, the VQLLE achieves better performance than the linear-algebra-based QLLE in reflecting the global manifold structure of the given data.   

\section{Conclusion}
\label{sec:conclusion}
In this paper, we have presented two quantum versions of the LLE algorithm which is a representative nonlinear dimensionality reduction algorithm. For the linear-algebra-based QLLE, we invoke the quantum $k$-NN algorithm to find out the $k$ nearest neighbors of the original high-dimensional data with quadratic speedup in data preprocessing. As to the main part of the algorithm, compared with the classical LLE, the linear-algebra-based QLLE can achieve exponential speedup in the number and dimension of the given data. In addition, we present the VQLLE utilizing a variational hybrid quantum-classical procedure to implement the dimensionality reduction on near term quantum devices. To evaluate the feasibility and performance of the QLLE algorithms proposed in our work, the numerical experiments are presented.

However, some open questions still need further study. Firstly, although the linear-algebra-based QLLE can achieve quantum speedup in theory, the results of the experiments on it shows that this algorithm requires high-depth quantum circuits and fully coherent evolution which are not achievable at present. In addition, the performance of the VQLLE depends largely on the specific design of the parameterized quantum circuits. Hence, how to design the structure of the circuits to achieve optimal performance still needs exploration.

\begin{acknowledgements}
We acknowledge support from the National Key R\&D Program of China, Grant No. 2018YFA0306703.
\end{acknowledgements}

\appendix

\section{Derivation of $W_{i}$}
\label{A}
According to the cost function Eq.~\eqref{eq:matrix_phi_W}, it is obvious that the Lagrange function
\begin{equation}\label{eq:L_{1}}
	\mathcal{L}_{1}(W_{i}, \mu) = W_i^T C_i W_i - \mu_{1} (1 - W_i^T \textbf{1}_{N}). 
\end{equation}
where $\mu_{1}$ represents the Lagrange multiplier and $i = 1, 2, \dots, N$.

Let's take the partial derivative of the Lagrange function $\mathcal{L}_{1}$ with respect to $W_{i}$ and set it to zero 
\begin{equation}\label{eq:partial_derivative_of_L_{1}}
	\frac{\partial \mathcal{L}_{1}}{\partial W_{i}} = 2C_{i}W_{i} + \mu_{1} \textbf{1}_{N} = 0
\end{equation}
resulting in 
\begin{equation}\label{eq:appendix_W_{i}_{1}}
	W_{i} = - \frac{\mu_{1}}{2} C_{i}^{-1} \textbf{1}_{N}.
\end{equation}

In addition, $W_{i}^{T} \textbf{1}_{N} = 1$ is equivalent to $\textbf{1}_{N}^{T} W_{i} = 1$. Thus, we have
\begin{equation}\label{eq:equivalent_equation}
	\textbf{1}_{N}^{T} (-\frac{\mu_{1}}{2} C_{i}^{-1} \textbf{1}_{N}) = 1
\end{equation}
and then
\begin{equation}\label{eq:mu_{1}}
	\mu_{1} = \frac{-2}{\textbf{1}_{N}^{T} C_{i}^{-1} \textbf{1}_{N}}.
\end{equation}

Finally,
\begin{equation}\label{eq:appendix_W_{i}}
	W_{i} = \frac{C_{i}^{-1} \textbf{1}_{N}}{\textbf{1}_{N}^{T} C_{i}^{-1} \textbf{1}_{N}}.
\end{equation}

\section{Derivation of $Y$}
\label{B}
According to Eq.~\eqref{eq:matrix_phi_y}, we can get the Lagrange function
\begin{equation}\label{eq:L_{2}}
	\mathcal{L}_{2}(Y, \Lambda, \beta) = \frac{1}{2} \Vert Y - YW \Vert^{2}_{F} - \frac{1}{2} \mathrm{tr}(\Lambda \left[ \frac{1}{N} YY^{T} - I_{d} \right]) - \textbf{1}_{N}^{T} Y^{T} \beta,
\end{equation}
where $\Lambda$ is a diagonal matrix whose diagonal elements are respectively the corresponding Lagrange multipliers and $\beta = [\beta_{1}, \dots, \beta_{d}]^{T}$ represents a $d$-dimensional vector whose elements are Lagrange multipliers.

We compute the partial derivative of the Lagrange function $\mathcal{L}_{2}$ with respect to $Y$ and set it to zero
\begin{equation}\label{eq:partial_derivative_of_L_{2}}
	\frac{\partial \mathcal{L}_{2}}{\partial Y} = Y(I_{N} - W)(I_{N} - W^{T}) - \frac{1}{N} \Lambda Y - \beta \textbf{1}^{T}_{N} = \textbf{0}.
\end{equation}

Then, we multiply $\textbf{1}_{N}$ on both sides of Eq.~\eqref{eq:partial_derivative_of_L_{2}} resulting in 
\begin{equation}\label{eq:after_multiplication}
	Y (I_{N} - W)(I_{N} - W^{T}) \textbf{1}_{N} - \frac{1}{N} \Lambda Y \textbf{1}_{N} - \beta \textbf{1}_{N}^{T} \textbf{1}_{N} = \textbf{0}.
\end{equation}

According to the constraint conditions of $W$ and $Y$, namely $W_{i}^{T} \textbf{1}_{N} = 1$ and $Y \textbf{1}_{N} = \textbf{0}$, the first and the second term of Eq.~\eqref{eq:after_multiplication} equal zero. Thus, $\beta = \textbf{0}$ because of $\textbf{1}_{N}^{T} \textbf{1}_{N} = N$.

Therefore,
\begin{equation}\label{eq:eigenequation_1_of_Y}
	YM = \frac{1}{N} \Lambda Y,
\end{equation}
where $M = (I_{N} - W)(I_{N} - W)^{T}$.

We diagonalize the matrix $\Lambda = UDU^T$, with $U$ is a orthogonal matrix and $D = \mathrm{diag}(d_1,d_2, \dots, d_d)$ with $d_{1} \leq d_{2} \leq \dots \leq d_{d}$. We let $\tilde{Y} = U^T Y$. Then Eq.~\eqref{eq:eigenequation_1_of_Y} can be transformed to
\begin{equation}\label{eq:eigenequation_2_of_Y}
	MY^{T} = \frac{1}{N} Y^{T}UDU^{T}
\end{equation}
which is equivalent to 
\begin{equation}\label{eq:eigenequation_3_of_Y}
	M \tilde{Y}^{T} = \frac{1}{N} \tilde{Y}^{T} D,
\end{equation}
where $\tilde{Y} = U^{T} Y$.

Eq.~\eqref{eq:eigenequation_3_of_Y} shows that the columns of $\tilde{Y}^T$(or the rows of $\tilde{Y}$) are eigenvectors of $M$. Because of $N^{-1}\tilde{Y}\tilde{Y}^T=I_d$, the norm of each column of $\tilde{Y}^T$ equals $N^{1/2}$. Hence, the columns of $N^{-1/2}\tilde{Y}^T$ are normalized eigenvectors of $M$.

%
%



\end{document}